# Imaging mechanical vibrations in suspended graphene sheets

D. Garcia-Sanchez,[1,2] A. M. van der Zande,[3] A. San Paulo,[2] B. Lassagne,[1,2] P. L. McEuen[3] and A. Bachtold,[*,1,2]

[1]CIN2 Barcelona, Campus UAB, E-08193 Bellaterra, Spain

[2]CNM-CSIC Barcelona, Campus UAB, E-08193 Bellaterra, Spain

[3]Cornell Center for Materials Research, Cornell University, Ithaca, NY 14853, USA.

* E-mail: adrian.bachtold@cnm.es

We carried out measurements on nanoelectromechanical systems based on multilayer graphene sheets suspended over trenches in silicon oxide. The motion of the suspended sheets was electrostatically driven at resonance using applied radio-frequency voltages. The mechanical vibrations were detected using a novel form of scanning probe microscopy, which allowed identification and spatial imaging of the shape of the mechanical eigenmodes. In as many as half the resonators measured, we observed a new class of exotic nanoscale vibration eigenmodes not predicted by the elastic beam theory, where the amplitude of vibration is maximum at the free edges. By modeling the suspended sheets with the finite element method, these edge eigenmodes are shown to be the result of non-uniform stress with remarkably large magnitudes (up to 1.5 GPa). This non-uniform stress, which arises from the way graphene is prepared by pressing or rubbing bulk graphite against another surface, should be taken into account in future studies on electronic and mechanical properties of graphene.

Graphene is a newly isolated material whose structure consists of a single layer of carbon atoms packed in a honeycomb crystal lattice[1,2]. Recently, stacks of graphene layers were suspended over a trench and clamped at the edges, obtaining a new type of nanoelectromechanical system (NEMS)[3]. Despite thicknesses all the way down to one atomic layer, these suspended stacks of graphene still maintain high crystalline order[4], resulting in a NEMS with extraordinarily small thickness, large surface area, low mass density, and high Young's modulus. Because of these excellent material properties, graphene NEMSs hold promise as very good detectors of mass, force and charge, and represent the ultimate limit of two dimensional NEMSs.

Previous work has shown suspended graphene sheets can be mechanically actuated, and the resonant frequencies are extracted using optical interferometry. However, this measurement technique can not identify what the individual vibrational eigenmodes are[3]. In this work, we directly image the spatial shape of the eigenmodes using a scanning force microscope (SFM). While the eigenmode shape can match predictions for doubly clamped beams typically discussed in elastic beam theory text books, we also observe new exotic eigenmodes in as many as half the suspended sheets measured. These exotic eigenmodes would be impossible to identify using more traditional measurement techniques, such as optical or capacitive detection, which depend on the average position of the resonator[3,5].

Few-layer suspended graphene sheets were obtained by mechanical exfoliation[6,7]. Highly ordered bulk graphite was pressed down onto a degenerately doped silicon wafer patterned with trenches etched into the oxide and with gold electrodes defined between the trenches. Suitable candidates for measurement were identified optically[8,9], looking for few-layer graphene sheets suspended over a trench and contacting at least one electrode. A scanning electron microscope image of a suspended graphene sheet used in these experiments is shown in Figure 1a.

The suspended graphene sheets were electrostatically actuated to become resonators by wiring them up as shown in Figure 1b. An oscillating radio frequency voltage $V_{RF}$ was applied between the backgate and the graphene, resulting in an oscillating electrostatic force at the same frequency:

$$F_{RF} = \frac{\partial C}{\partial z}(V_{DC} - \phi)V_{RF} \qquad (1)$$

where $V_{DC}$ is the DC voltage applied to the gate, $\phi$ is the contact potential between the resonator and the gate, and $\frac{\partial C}{\partial z}$ is the spatial derivative of the capacitance between the resonator and the gate.

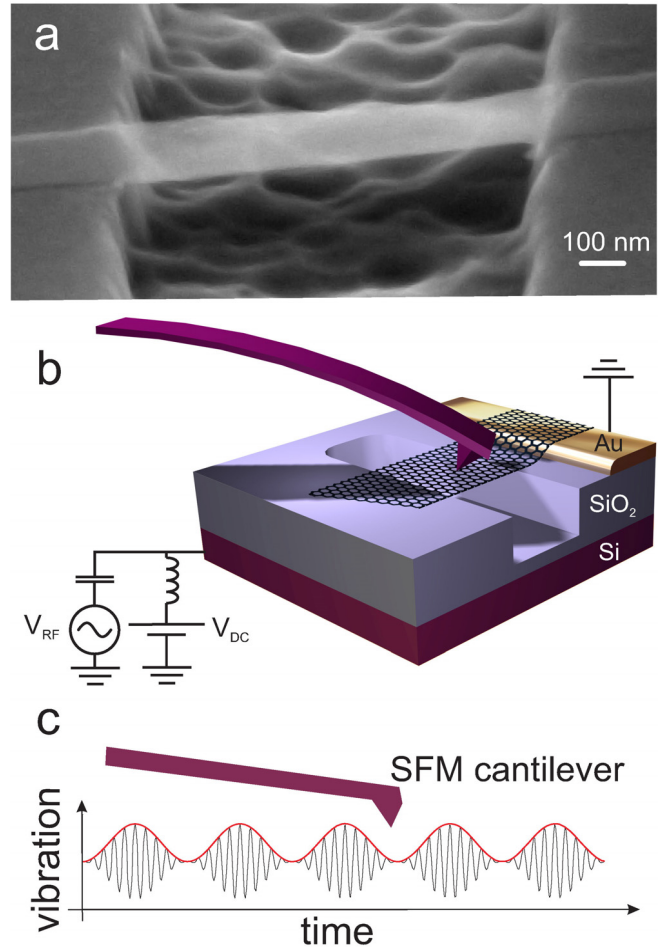

**Figure 1** Device and experimental setup. **a,** A scanning electron microscope image of a suspended graphene resonator. **b,** Schematic of the resonator together with the SFM cantilever. **c** Motion of the suspended graphene sheet as a function of time. A high-frequency term at $f_{RF}$ is matched to the resonance frequency of the graphene, and the resulting oscillation is modulated at $f_{mod}$.

The mechanical vibration of the resonator was detected using a recently reported SFM technique, which allowed the measurement of the resonance frequency as well as the shape of the eigenmode[10]. This technique is particularly suitable for the detection of low amplitude vibrations, which here are typically of the order of 0.1 nm (see Supporting Information). As shown in Figure 1b, a SFM cantilever scans over the surface of the suspended sheet, while the resonator is being driven. The frequency $f_{RF}$ of the driving voltage $V_{RF}$ is set at (or close to) the resonance frequency of the resonator. In addition, $V_{RF}$ is modulated at $f_{mod}$, $(1-\cos(2\pi f_{mod}t))\cos(2\pi f_{RF}t)$, so the resonator vibrations are sequentially turned on and off at $f_{mod}$ (Fig. 1c). While the SFM cantilever cannot follow the rapid oscillations at $f_{RF}$, it can detect the height difference between the on and off states of the modulation envelope. In order to enhance the detection sensitivity to the vibrations, $f_{mod}$ is matched to the resonance frequency of the first eigenmode of the SFM cantilever. At the same time, the topography is obtained with the second eigenmode of the cantilever to keep the tip at a constant height above the surface. Further details on the technique can be found in Supporting Information.

| l(μm) | t(nm) | w(μm) min | w(μm) max | $f_1$(MHz) | $f_2$(MHz) | Mode |
|---|---|---|---|---|---|---|
| 2.9 | 1 | 0.1 | 0.8 | 18 | - | Beam |
| 2.7 | 3 | 0.8 | 1.8 | 45 | - | Beam |
| 4.4 | 6 | 0.5 | 0.8 | 33 | - | Beam |
| 1.8 | 10 | 0.2 | 0.6 | 37 | - | Beam |
| 2.8 | 11 | 0.3 | 0.5 | 31 | - | Beam |
| 2.9 | 20 | 0.6 | 1.0 | 57 | - | Beam |
| 4.2 | 4 | 1.3 | 1.5 | 25 | 59 | Edge |
| 2.8 | 5 | 0.8 | 1.0 | 47 | - | Edge |
| 3.5 | 6 | 1.0 | 1.4 | 33 | 70 | Edge |
| 2.8 | 6 | 0.5 | 0.8 | 53 | 85 | Edge |
| 2.9 | 10 | 0.7 | 1.5 | 26 | - | Edge |

**Table 1.** Resonator characteristics. *l* is the length of the resonator, *t* the thickness, *w* the width, and $f_1$ ($f_2$) the resonance frequency of the first (second) eigenmode. *w* can significantly vary along the resonator, so we report the minimum and the maximum width.

Table 1 summarizes the dimensional characteristics of the suspended sheets that we have studied, such as the thickness *t*, length *l*, and width *w*, as well as the resonance frequencies. The measured quality factors are low (between 2 and 30). This low Q is attributed to energy dissipation to air molecules since the SFM technique is operated at atmospheric conditions[10,11]. The shape of the eigenmodes does not change as the frequency is swept (within a same resonance peak of the amplitude versus frequency). All measurements are performed in the linear amplitude response regime. See Supporting Information for further details.

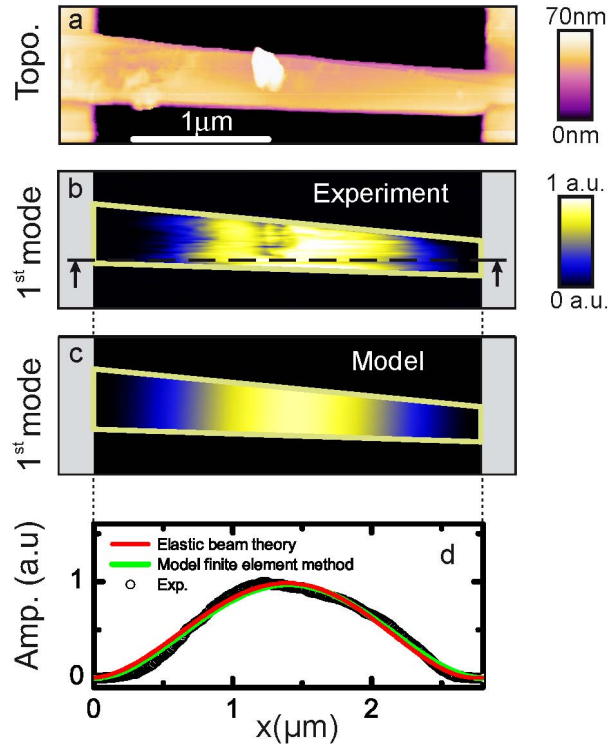

**Figure 2** Graphene resonator with no buckling. **a,** Measured topography. The image reveals irregularities on the surface of the resonator, presumably due to contamination or bulk graphite residue. $t = 11$ nm, $l = 2.8$ um, $w_{min} = 0.3$ um and $w_{max} = 0.5$ um. **b,** Measured shape of the eigenmode at 31 MHz (raw data). $V_{DC}$-$\phi$ = 3V and $V_{RF}$ = 60mV. The amplitude of vibration is in arbitrary units. **c,** Shape of the eigenmode at 31 MHz obtained using FEM simulations without any stress. **d,** Eigenmode along the line indicated in Fig 2b.

The SFM technique yields high resolution images of the shape of the vibration eigenmodes, since the scanning SFM tip measures the amplitude of vibration as a function of position. We find two distinct types of eigenmodes in the suspended graphene resonators. Figures 2a and 3a show the height topography images of two different suspended graphene sheets with thicknesses of 11 and 6 nm. Figures 2b and 3b,c show the corresponding eigenmodes with respective resonance frequencies 31, 53, and 85 MHz. For some resonators, the amplitude of vibration remains uniform across the width of the sheet similar to standard beam resonators (Fig. 2b), while, for others, the vibrations are the largest in amplitude along one of the free edges (Fig. 3b and c). We call the former modes "beam modes", and the latter "edge modes".

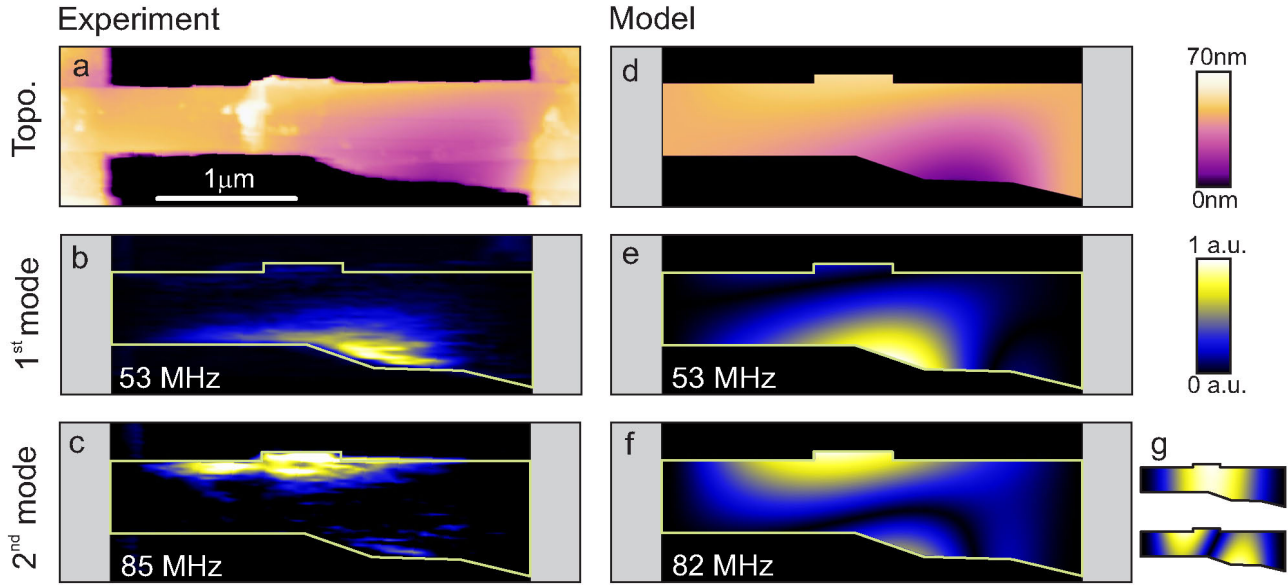

**Figure 3** Graphene resonator with local buckling. **a,** Measured topography. $t$ = 6nm, $l$ = 2.8um, $w_{min}$ = 0.5 um and $w_{max}$ = 0.8 um. The maximum out of plane displacement of the buckling is 37 nm. **b-c,** Shape of the first and the second eigenmodes (raw data). $V_{DC}$-$\phi$ = 3V and $V_{RF}$ = 40mV. The amplitude of vibration is in arbitrary units. **d,** Topography obtained using FEM simulations on a stressed graphene sheet. The maximum displacement is 36 nm. See the text for the boundary conditions. **e-f,** Shape of the first and the second eigenmodes using FEM simulations. **g,** Shape of the two first eigenmodes using FEM simulations without any stress. The resonance frequencies are 17 and 46 MHz.

Let us first compare the beam modes to predictions from the elastic beam theory, which describes the dynamics of a linear resonator with a uniform cross-section[12]. A beam under weak uniform tension $T$ is predicted to have a fundamental mode resonance frequency of

$$f_1 = \frac{1.028}{l^2}\sqrt{\frac{Et^2}{\rho}} + T\frac{0.154}{\sqrt{\rho E w^2 t^4}} \qquad (2)$$

where the shape of the first eigenmode along the beam axis $x$ is predicted to be

$$z_1 = \left[\cos\left(\frac{4.73x}{l}\right) - \cosh\left(\frac{4.73x}{l}\right)\right] - 0.98\left[\sin\left(\frac{4.73x}{l}\right) - \sinh\left(\frac{4.73x}{l}\right)\right] \quad (3)$$

where the Young's Modulus is $E = 1$ TPa, and the mass density is $\rho = 2200$ kg m$^{-3}$. Taking the values $l = 2.8$ μm and $t = 11$ nm for the graphene sheet shown in Figure 2 and assuming $T = 0$, equation 2 predicts a resonance frequency $f_l = 31$ MHz in excellent agreement with the experimentally measured value. In addition, we find that equation 3 qualitatively describes the measured eigenmode shape (as shown in Figure 2d). Note, however, that equations 2 and 3 are calculated when the resonator is perpendicular to the clamping edge and has a constant width along the beam axis. The topography image of the suspended sheet shows that this is not the case (Fig. 2a). We have developed a model based on the finite element method (FEM) to take these complications into account (see Supporting Information). Figures 2c and d show the predicted eigenmode shape of the resonator according to FEM. The predicted eigenmode has resonance frequency of 31MHz, in excellent agreement with the measured value. The shape agrees qualitatively with the measurements and remains very similar to the predictions of the elastic beam theory.

Not all suspended graphene sheets display such conventional eigenmodes. The edge modes shown in Figures 3b and c are completely unpredicted by standard elastic beam theory. However, the topography image of the suspended sheet in Figure 3a shows that the resonator is buckled out of plane at one edge. This local buckling is measured to have a maximum out of plane displacement of 37 nm and suggests the presence of non-uniform stress in the resonator. The edge modes are frequently, but not always, observed in resonators for which the suspended sheet displays local buckling.

To understand the relationship between local buckling and the edge modes, we calculate the effect of strain with simulations based on the finite element method (see Supporting Information). The strain is introduced by imposing an in-plane stretch and in-plane rotation to the suspended sheet at the clamping edges. In Figure 3d, we impose an in-plane stretch of 1.5 nm and in-plane rotation of -0.2 degrees about a pivot point at the top right of the resonator. This strain results in local buckling in the lower edge with a maximum out of plane displacement of 36 nm, consistent with measurement. Figures 3e and f show the predicted eigenmodes for the buckled sheet. Both the resonance frequencies and shapes are in reasonable agreement with measurements. For comparison, simulations carried out without any stress result in conventional beam modes as shown in Figure 3g.

This good agreement overall allows us to use the simulation to evaluate the stress in the resonator. Stress is described by a 3x3 tensor that varies over the volume of the resonator[13]:

$$\sigma_{ij} = \begin{bmatrix} \sigma_x & \tau_{xy} & \tau_{xz} \\ \tau_{yx} & \sigma_y & \tau_{yz} \\ \tau_{zx} & \tau_{zy} & \sigma_z \end{bmatrix} \quad (4)$$

where $\sigma_i$ is the stress along axis $i$ and $\tau_{ij}$ is the shearing stress in plane $ij$. A suitable transformation of the direction of the coordinate axes through rotation allows the components of the shearing stress $\tau_{ij}$ to vanish. The result is a diagonal matrix with 3 components: $\sigma_1$, $\sigma_2$ and $\sigma_3$, where $\sigma_1 > \sigma_2, \sigma_3$ and $\sigma_1$ is known as the principal stress. Note that this transformation is made for each point of the resonator. The maximum value of the principle stress before a material breaks is used as the failure criteria for crystalline materials[13].

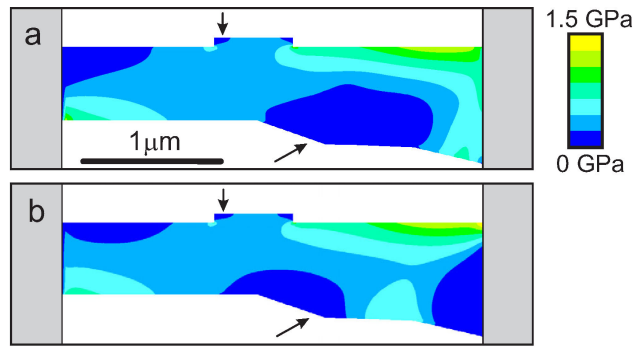

**Figure 4** Calculated principal stress $\sigma_1$ for the suspended graphene sheet in Figure 3. The colour scale shows the stress in the resonator, ranging from 0 to 1.5 GPa. **a,** Top surface. **b,** Bottom surface. Arrows indicate regions with low stress where the amplitude of the eigenmodes is maximum.

Figure 4 shows the spatial distribution of $\sigma_1$ over the top and the bottom surfaces of the suspended sheet from Figure 3. In between these two boundary surfaces $\sigma_1$ varies continuously (not shown). The maximum stress in the suspended sheet is very high, about 1.5 GPa. For comparison, it has been shown that 1020 Steel breaks at 690 MPa[13], MWNTs between 11 and 63 GPa[14], and SWNTs between 13 and 52 GPa[15]. We attribute this high stress to be the consequence of the macroscopic uncontrolled forces that are applied during the mechanical exfoliation process step. Part of the resulting stress remains after the applied forces are released due to the large van der Waals interaction that holds the graphene to the $SiO_2$ surface.

Comparing Figures 3 and 4, it appears that the distribution of stress and the resonance properties are closely related. The vibration amplitude is larger in regions of lower stress. In addition, there is a correlation between the resonance modes and regions of increasing stress: the fundamental eigenmode resonates at the lower free edge where the stress is the lowest, and the second eigenmode resonates at the upper edge where the stress is larger. The underlying physical mechanism is the same as for a beam under uniform tension for which the resonance frequency increases as the tension is raised[16].

To conclude, we have imaged the eigenmode shape of graphene resonators. For some resonators, we have found a new class of nanoscale eigenmodes where the vibrations are maximum in amplitude not at the center of the beam, but at the free edges. Simulations based on the finite element method indicate that these eigenmodes are the result of the high non-uniform stress present in the resonator. The shape of these exotic eigenmodes and the corresponding stress must be taken into account in future experiments and applications, such as for the determination of the Young's modulus[17,18] and the accurate calibration of mass, force, or charge sensing[3,19,20]. It could also be possible to manipulate the eigenmode shape by varying the strain during measurements via electrostatic tuning[3], varying the pressure difference across sealed membranes, or displacing the clamping edges. This would allow one to activate mechanical vibrations in localized areas for multiple-target sensing applications. It is also important to realize that stress can be present in graphene sheets regardless of whether they are suspended or not. This stress needs to be taken into account when engineering the band gap of graphene ribbons[21,22,23], in measuring the amplitude and wavelength of ripples[24,25], or in estimating the effective magnetic field in quantum electron interference experiments[26,27].

**Acknowledgement** We thank Scott Verbridge and Jeevak Parpia for help with sample characterization. The research has been supported by a EURYI grant and FP6-IST-021285-2 and by the NSF through the Cornell Center for Materials Research. Sample fabrication was performed at the Cornell Nanoscale Science and Technology Facility, a National Nanotechnology Infrastructure Network node, funded by NSF.

**Supporting Information Available:** A detailed description of fabrication technique, a description of the FEM model, a comparison of the FEM model to analytical predictions for a beam under tension[28] as well as recent simulations on nanotubes with slack[29], and a detailed description of the SFM technique to detect mechanical vibrations are available free of charge via the Internet at http://pubs.acs.org.

1 Castro Neto, A.; Guinea, F.; Peres, N. M. *Phys. World* **2006,** *19,* 33.

2 A. K. Geim and K. S. Novoselov, *Nat. Mater.* **2007,** 6, 183.

3 Bunch, J.S.; et al. *Science* **2007,** *315,* 490.

4 Meyer, J.C.; et al. *Nature* **2007,** *446,* 60.

5 Sazonova, V.; et al. *Nature* **2004,** *431*, 284.

6 Novoselov, K.S.; et al *Science* **2004,** *306,* 666.

7 Novoselov K.S.; et al. *Proc. Natl. Acad. Sci. U.S.A.* **2005,** *102,* 10451.

8 Blake, P.; et al. *App. Phys. Lett.* **2007,** *91,* 063124.

9 Abergel, D.S.L.; Russell, A.; Fal'ko, V.I. *App. Phys. Lett.* **2007,** *91,* 063125.

10 Garcia-Sanchez, D.; et al. *Phys. Rev. Lett.* **2007,** *99,* 085501.

11 Ekinci, K.L.; Roukes, M.L. *Rev. Sci. Instrum*. **2005,** *76,* 061101.

12 Weaver, W.; Timoshenko, S.P.; Young, D.H. *Vibrations problems in engineering*. Wiley, New York, 1990.

13 Ford, H.; Alexander, J.M. *Advanced Mechanics of Materials*. Longmans, London, 1963.

14 Yu, MF.; et al. *Science* **2000,** *287,* 637.

15 Yu, MF.; et al. *Phys. Rev. Lett.* **2000,** *84,* 5552.

16 Verbridge, S.S.; et al. *Nano Lett*. **2007,** *7,* 1728.

17 Frank, I.W.; Tanenbaum, D. M.; van der Zande, A.M.; McEuen P.L. *J. of Vac. Sci. And Tech. B,* **2007,** *25,* 2558.

18  Poot, M; van der Zant, H.S.J. *App. Phys. Lett.,* **2007,** 92, 063111.

19 Craighead, H. *Nature Nanotech*. **2007,** *2,* 18.

20 Tamayo, J.; Ramos, D.; Mertens, J.; Calleja, M. *Appl. Phys. Lett*. **2006,** *89,* 224104.

21 Fiori, G.; Iannaccone, G. *IEEE Electron Device Lett.* **2007,** *28,* 760.

22 Yang, L.; Anantram, M.P.; Han, J.; Lu, J.P. *Phys. Rev. B* **1999,** *60,* 13874.

23 Yang, L.; Han, J. *Phys. Rev. Lett.* **2000,** *85,* 154.

24 Fasolino, A.; Los, J.H.; Katsnelson, M.I. *Nature Materials* **2007,** 6, 858.

25 Morozov, S.V. ; et al. *Phys. Rev. Lett.* **2008,** 100*,* 016602.

26 Morozov, S.V. et al. *Phys. Rev. Lett.* **2006,** *97,* 016801.

27 Morpurgo, A.F.; Guinea F. *Phys. Rev. Lett.* **2006,** *97,* 196804.

28 Buks, E.; Roukes, M.L. *Phys. Rev B* **2001,** *63,* 033402.

29 Üstunel,; H., Roundy, D.; Arias. T.A. *Nano Lett.* **2005,** *5,* 523.

Supporting Information
# Imaging mechanical vibrations in suspended graphene sheets
D. Garcia-Sanchez, A. M. van der Zande, A. San Paulo, B. Lassagne, P. L. McEuen and A. Bachtold

**A - Sample fabrication**

Suspended graphene sheets are fabricated by mechanical exfoliation. A freshly cleaved piece of Kish graphite (Toshiba Ceramics) is rubbed onto a degenerately doped silicon wafer with 290 nm $SiO_2$ grown by plasma enhanced chemical vapor deposition. Before depositing the graphene, the wafer is patterned with trenches using photolithography and plasma etching. The trenches are millimeter long, 0.5 – 10 μm wide, and 250 nm deep. Electrodes defined by photolithography between the trenches are deposited using electron-beam evaporation and consist of 5 nm Cr and 35 nm Au.

**B- Description of the FEM model**

To model the shape of the eigenmodes, we have developed a simulation based on finite element methods (FEM) using ANSYS. The first step of the simulation is to account for the buckling of the suspended sheet by finding the adequate boundary conditions at the clamping edges. To do this, we hold one clamping edge of the suspended region fixed, and impose an in plane displacement to the other clamping edge. Specifically, the displacement of this edge consists of a translation and a rotation within the undeformed resonator surface. Since the resulting out of plane displacement can be large, calculations are carried out taking into account geometric non-linear deformations[1]. To ensure that the buckling goes in the desired direction, we apply an out of plane perturbative force, which is then cleared at the end of the calculations. We make the assumption that the mechanical properties of the resonator are isotropic with 1TPa for the Young's modulus and 0.17 for the Poisson ratio[2]. The exact value of the Poisson ratio has little effect on the output of the calculations. In the second step of the simulation, a modal analysis is performed to determine the resonance frequency and the eigenmode shape of the deformed resonator. Here, the modal analysis is carried out in the linear regime because the amplitude of the vibration is small. To check that this simulation is free of errors, it has been successfully compared to analytical predictions for a beam under tension[3]. The simulation also reproduces recent calculations on nanotubes with slack[4]. See sections C and D.

**C- Comparison between the FEM model and analytical expressions for resonators under tension.**

The resonance frequency for a beam under weak uniform tension ($T << EI/l^2$) is[5]

$$f_1 = \frac{22.4}{2\pi l^2}\sqrt{\frac{EI}{\rho wt}} + \frac{0.28}{2\pi}T\sqrt{\frac{1}{\rho wtEI}} \qquad (1)$$

We take the length $l$ = 500 nm, the width $w$ = 20 nm, the thickness $t$ = 3.5 Å, the density $\rho$ = 2200 kg/m$^3$, and the Young's Modulus $E$ = 1TPa. The bending moment of a rigid beam is $I = wt^3/12$.

For high tension ($T>>EI/l^2$) the frequency can be expressed as[3]

$$f_1 = \frac{1}{2l}\sqrt{\frac{T}{\rho wt}}\left[1 + 2\xi + (4 + \pi^2/2)\xi^2\right] \qquad (2)$$

with $\xi^2 = Etw^3/12Tl^2$.

Figure S1 shows the resonance frequency as a function of tension using the FEM model and the above expressions. There is a good agreement between the theory and the FEM model.

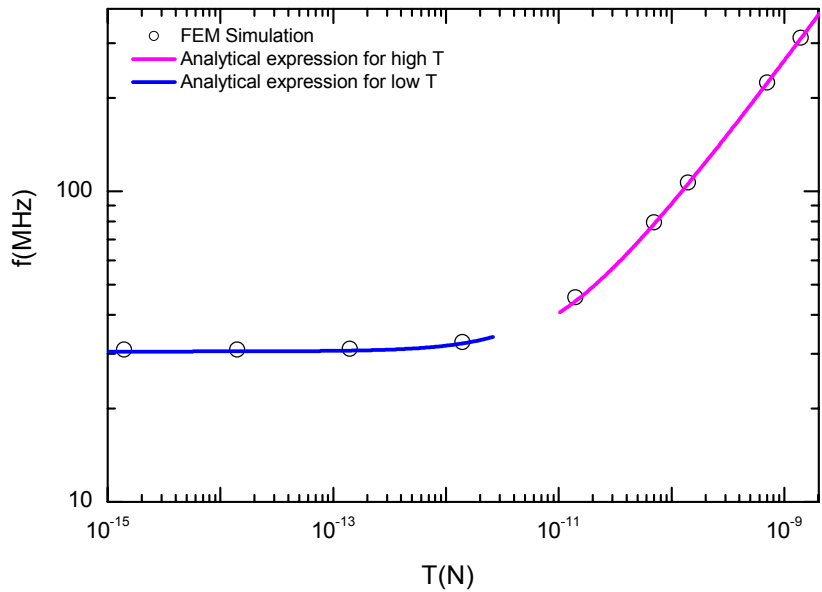

**Figure S1** Resonance frequency as a function of tension for the first eigenmode of a graphene resonator under tension.

**D- Comparison between the FEM model and previous simulations on buckled SWNTs.**

Previous work[4] has reported numerical studies on SWNT resonators that are buckled (slack). We compare this work to the FEM model that we have developed. For this, we use the same geometry and the same physical characteristics as in reference 4. The resonator is a doubly clamped rod with $l$=1.75um, $d$=2nm, $E$=2.18TPa, $\rho$=2992kg/m3 and a slack of 0,3%. The slack is defined as the ratio of the excess length of the tube to the distance between the clamping points. Figure S2 shows the resonance frequency for different eigenmodes obtained with the two simulation techniques. A good agreement is obtained.

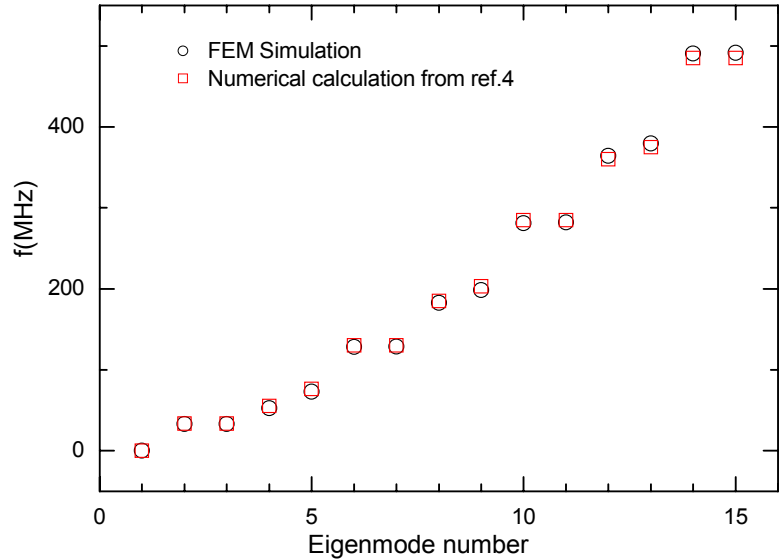

**Figure S2** Resonance frequency for different eigenmodes of a SWNT resonator with 0.3% of slack.

**Conclusions of sections C and D.** The FEM model shows a good agreement with established analytical expressions for beam resonators under tension and previous numerical calculations on buckled SWNTs. Note that the FEM model that we have developed can go beyond these cases. The model can be applied to resonators with any arbitrary geometry and any arbitrary stressed state.

**E- SFM detection of mechanical vibrations.**

We have developed a technique based on scanning force microscopy (SFM) to detect the mechanical vibrations of nanotube and graphene resonators[6]. The resonator motion is electrostatically actuated with an oscillating voltage applied on a gate electrode. The frequency $f_{RF}$ of the driving voltage $V_{RF}$ is set at (or close to) the resonance frequency of the resonator. In addition, $V_{RF}$ is modulated at $f_{mod}$, $(1-\cos(2\pi f_{mod}t))\cos(2\pi f_{RF}t)$. While the SFM cantilever cannot follow the rapid oscillations at $f_{RF}$, it can detect the modulation envelope.

The topography imaging is obtained in tapping mode using the second eigenmode of the SFM cantilever. The vibrations are detected with the first eigenmode of the SFM cantilever. Figure S3 shows that the signal of the vibrations is significantly enhanced when $f_{mod}$ is matching the resonance frequency $f_{tip}$ of the first eigenmode of the SFM cantilever. As a result, measurements are carried out with $f_{mod} = f_{tip}$.

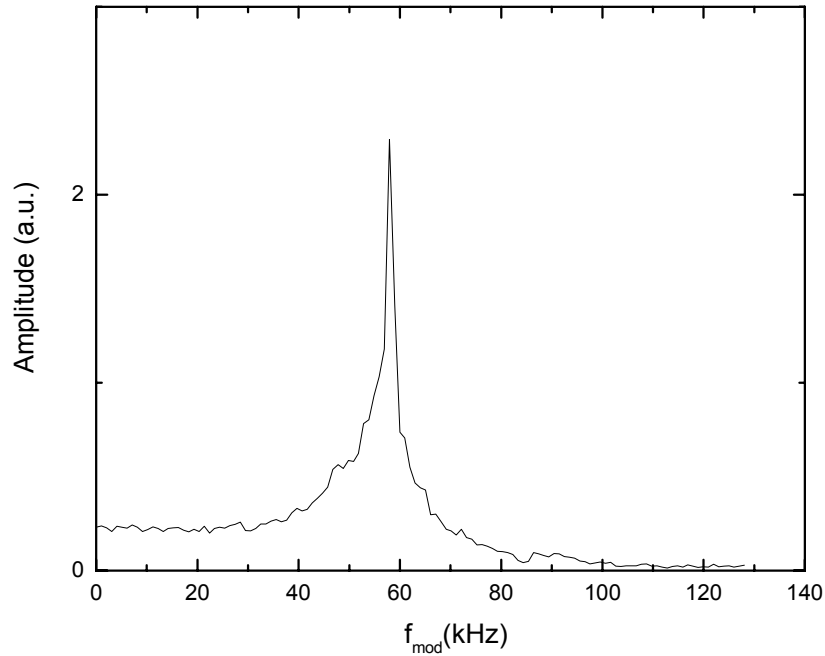

**Figure S3** Detected response of the vibration of a nanotube driven at resonance as a function of $f_{mod}$. The frequency of the first eigenmode of the SFM cantilever is 58 kHz. Nanotube resonance frequency $f_{RF}$ is 153 MHz. Measurements are taken at the nanotube position where the vibration amplitude is maximum.

Graphene resonators show a lorentzian response to the rf drive frequency. Figure S4 a shows the frequency response of a graphene resonator measured with the SFM technique. For comparison, Fig. S4 b shows the frequency response of the same resonator measured using optical interferometry[7]. The resonance frequencies are very similar for both techniques. However, the quality factor measured with the SFM technique is much lower due to energy dissipation to air, as the SFM technique is operated at atmospheric conditions, while the optical interferometry is performed in vacuum. Note that the low Q is not attributed to the disturbance of the SFM tip[6]. Indeed, we have noticed no change in the quality factor as the amplitude set point of the SFM cantilever is reduced by 3%–5% from the limit of cantilever retraction, which corresponds to the enhancement of the resonator-tip interaction.

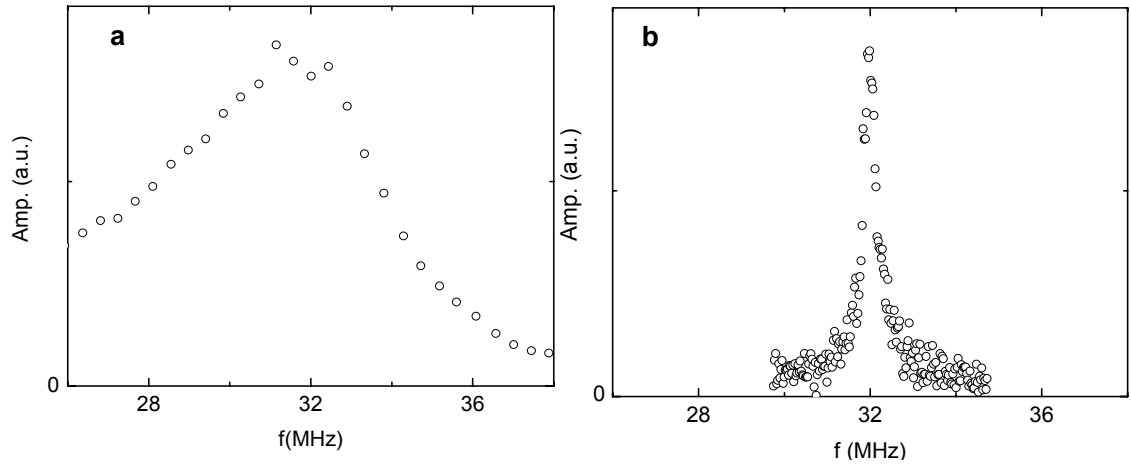

**Figure S4 a** Resonance peak of the fundamental mode of the graphene resonator shown in Figure 2 of the paper. The measurement is carried out using the SFM technique in air. The resonance is found at ~31 MHz with the quality factor Q = 5. Measurements are taken at the position where the vibration amplitude is maximum. **b** Resonance peak measured optically with a pressure of < $10^{-6}$ torr. The resonance is found at 32 MHz with Q = 64.

As shown in Eq. 1 of the paper, the radio frequency force $F_{RF}$ on the suspended sheet is a linear function of the offset voltage $V_{DC}$ and the radio frequency voltage $V_{RF}$. Figure S5 shows vibration amplitude as a function of $V_{DC}$ for an edge eigenmode of a graphene resonator. We find that the vibration amplitude is a linear function of the DC voltage and thus of the force. By operating in this regime, we ensure that the resonators are operating in the linear response regime, and the exotic edge eigenmodes are not a result of non-linear effects.

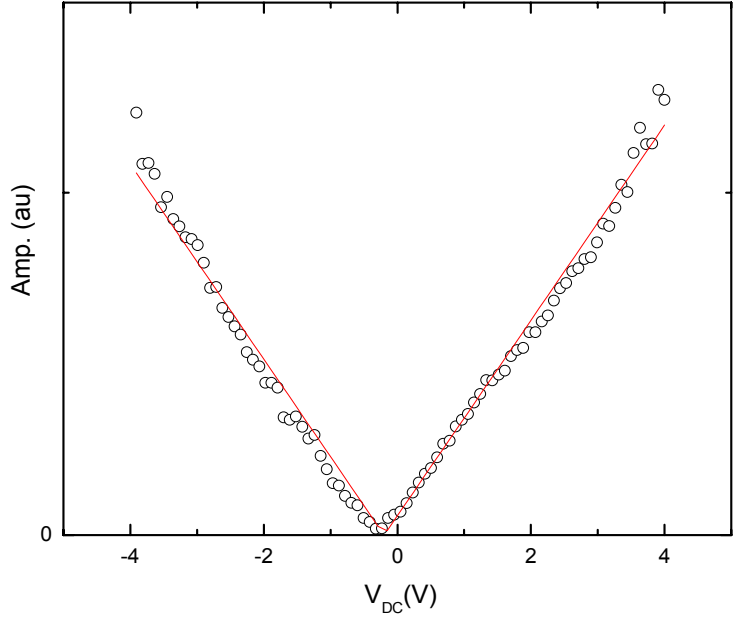

**Figure S5** Vibration amplitude as a function of $V_{DC}$ at $f_1$ = 33 MHz and $V_{RF}$ = 100 mV for an edge eigenmode of a graphene resonator. The dimensions of the suspended sheet are $t$ = 6 nm and $l$ = 3.5 μm. The measurements are obtained at the position of the graphene sheet where the amplitude is maximum

**F- Estimation of the vibration amplitude of graphene resonators.**

The amplitude of the vibration can be estimated when considering beams that neither have slack nor tension. The eigenfunctions $U_n(x)$ are obtained from[6,8]:

$$\rho A \frac{\partial U_{beam}}{\partial t^2} + EI \frac{\partial U_{beam}}{\partial x^4} = 0 \qquad (4)$$

The displacement of the resonator can be expanded in terms of $U_n(x)$,

$$z_{beam} = \left| \sum \alpha_n U_n \exp(-i 2\pi f_{RF} t) \right| \qquad (5)$$

where

$$\alpha_n = \frac{1}{4\pi^2 A \rho L^3} \frac{1}{f_n^2 - f_{RF}^2 - i f_n^2 / Q_n} \int_0^l U_n(x) F_{RF}(x) dx \qquad (6)$$

with $f_n$ the eigenfrequencies, $Q_n$ the quality factor for each eigenmode and $F_{RF} = \partial C(x)/\partial(z)(V_{DC} - \phi)V_{RF}$. C(x) is the capacitance per unit of length and is given by $C(x) = \varepsilon_0 w / z$ with w the width of the resonator and z the separation between the resonator and the gate. We have estimated that the maximum amplitude of the graphene resonator in Figure 2 of the paper is 0.1 nm. For this, we have used Q = 5, $V_{DC} - \phi$ = 3V, and $V_{RF}$ = 60 mV.

1 Zienkiewick, O.C.; Taylor, R.L. *The Finite Element Method 5th edition*. Butterworth-Heinemann; Oxford, 2000.

2 Popov, V.N.; Van Doren, V.E.; Balkanski M. Elastic properties of single-walled carbon nanotubes. *Phys. Rev. B* **2000,** *61,* 3078.

3 E. Buks, M.L. Roukes, Phys. Rev. B **2001,** *63,* 033402.

4 Üstunel,; H., Roundy, D.; Arias. T.A. *Nano Lett.* **2005,** *5,* 523.

5 S. Sapmaz, *et al.*, Phys.Rev B **2003,** *67,* 235414.

6 Garcia-Sanchez, D.; et al. *Phys. Rev. Lett.* **2007,** *99,* 085501.

7 Bunch, J.S.; et al. *Science* **2007,** *315,* 490.

8 A. N. Cleland., *Fondations of Nanomechanics* (Springer, Berlin, 2003).